\documentclass[9pt]{IEEEtran}
\IEEEoverridecommandlockouts
\usepackage{amsmath,amssymb,amsfonts}
\usepackage{algorithmic}
\usepackage{graphicx}
\usepackage{booktabs}
\usepackage{microtype}
\usepackage{textcomp}
\usepackage{xcolor}
\usepackage[square,numbers]{natbib}
\def\BibTeX{{\rm B\kern-.05em{\sc i\kern-.025em b}\kern-.08em
    T\kern-.1667em\lower.7ex\hbox{E}\kern-.125emX}}
\begin{document}

\title{Hierarchical Symbolic Pop Music Generation with Graph Neural Networks}

% \author{\IEEEauthorblockN{1\textsuperscript{st} Wen Qing LIM}
% \IEEEauthorblockA{\textit{Center for Digital Music} \\
% \textit{Queen Mary University of London}\\
% London, United Kingdom \\
% w.lim@se23.qmul.ac.uk}
% \and
% \IEEEauthorblockN{2\textsuperscript{nd} Huan ZHANG}
% \IEEEauthorblockA{\textit{Center for Digital Music} \\
% \textit{Queen Mary University of London}\\
% London, United Kingdom \\
% huan.zhang@qmul.ac.uk}
% \and
% \IEEEauthorblockN{3\textsuperscript{rd} Jinhua LIANG}
% \IEEEauthorblockA{\textit{Center for Digital Music} \\
% \textit{Queen Mary University of London}\\
% London, United Kingdom \\
% jinhua.liang@qmul.ac.uk}
% }

% \author{
% \name{Wen Qing Lim \qquad Jinhua Liang \qquad  Huan Zhang}\\
% \address{\textit{ Queen Mary University of London}, Centre for Digital Music, London, United Kingdom } \\
% w.lim@se23.qmul.ac.uk, jinhua.liang@qmul.ac.uk, huan.zhang@qmul.ac.uk
% }

\author{
Wen Qing Lim \qquad Jinhua Liang \qquad  Huan Zhang\\
\textit{ Queen Mary University of London}, Centre for Digital Music, London, United Kingdom \\
w.lim@se23.qmul.ac.uk, jinhua.liang@qmul.ac.uk, huan.zhang@qmul.ac.uk
}

\maketitle

\begin{abstract}
Music is inherently made up of complex structures, and representing them as graphs helps to capture multiple levels of relationships. While music generation has been explored using various deep generation techniques, research on graph-related music generation is sparse. Earlier graph-based music generation worked only on generating melodies, and recent works to generate polyphonic music  do not account for longer-term structure.  In this paper, we explore a multi-graph approach to represent both the rhythmic patterns and phrase structure of Chinese pop music. Consequently, we propose a two-step approach that aims to generate polyphonic music with coherent rhythm and long-term structure. We train two Variational Auto-Encoder networks - one on a MIDI dataset to generate 4-bar phrases, and another on song structure labels to generate full song structure. Our work shows that the models are able to learn most of the structural nuances in the training dataset, including chord and pitch frequency distributions, and phrase attributes.  
\end{abstract}

\begin{IEEEkeywords}
Music generation, graph neural network, symbolic music generation
\end{IEEEkeywords}

\section{Introduction}

% [Why Graph representation?] 
Music, an intricate tapestry of rhythm, melody, and harmony, is inherently structured in a way that lends itself well to graph representations. At a higher level, music can viewed as a progression of sections and phrases, while at a more granular level, it is made up of complex interplay of notes, chords, and rhythms.  Both levels of music structure can intuitively be captured in graphs. Meanwhile, symbolic music formats (e.g. MIDI, XML) contains structured information about pitch, duration, onset time, and tracks \cite{Cancino-Chacon2022PartituraProcessing}. This makes symbolic music an ideal format to be represented as graphs \cite{Zhang2023SymbolicEvaluation, Karystinaios2022CadenceNetworks}.

% [Current state of symbolic music generation and its challenges] 
Music generation has seen rapid advancements, particularly with the increasing popularity of text-to-audio models \cite{copet2024simplecontrollablemusicgeneration, Liu2023AudioLDMModels, Liang2024a} that create complete songs from text prompts. While raw audio generation has the advantage of producing high-fidelity music that can be used for creative purposes, it lacks the level of customization that is needed for music composition and arrangement. Symbolic music allows for more precise control over musical elements like melody, harmony, and rhythm. 

Symbolic music generation has made great progress in capturing inter-track relationships like chords and rhythm \cite{surveysymbolicmugen10.1145/3597493}, but still struggle with long term structure. \cite{Dannenberg2020, Yin2023DeepGeneration, Ma2024FoundationSurvey} While work like MELONS \cite{zou2021melonsgeneratingmelodylongterm} and PopMNet \cite{popmnetWU2020103303} have found success in generating longer forms of music using graph representations, they work only on melody generation instead of polyphonic music. On the other hand, work like Polyphemus \cite{cosenza2023graphbasedpolyphonicmultitrackmusic} have successfully generated structured polyphonic music using graph representations, but lack the ability to extend beyond a few bars. An intuitive question arises: can we derive a model that creates both polyphonic and long-term music?

% [What this paper sets out to achieve] 
In this paper, we aim to generate polyphonic symbolic music with long-term structure with a hierarchical two-stage framework. First, we train two separate Variational Auto-Encoder (VAE)s to encode and decode graph representations of song structure and phrases respectively. Each VAE encodes its input as an embedding in a latent space. Using the trained VAEs, we then generate new song structure and phrases by passing a random latent embedding through the model's decoder. Combining the two models, the generated phrases are interpolated according to the song structure to form sequences of coherent phrases that make up a song. 

The contributions of the paper are: 
\begin{enumerate}
    \item We proposed a novel graph representation that captures song structure based on POP909 structural annotation, including the relationships between phrases, and the phrase attributes..
    \item We explored the use of VAE graph network for song structure generation based on the proposed structure graph. We then employed a two-stage approach that combines 4-bar phrase prediction and structure prediction.
    \item We conducted in-depth evaluation analysis regarding  the structural aspect that follows existing literature, demonstrating that our generated songs have structure that is closer to human-composed songs.
\end{enumerate}
\section{Related Work}
The problem we have laid out concerns three areas of research: (A) Symbolic music generation, (B) Graph representations of music, and (C) Graph generation. Each will be discussed in their respective sections below.

\subsection{Symbolic Music Generation}

One of the challenges in symbolic music generation lies in capturing the intricacies and multi-faceted qualities \cite{Ma2024FoundationSurvey, Zhang2024DExterModels} of human-composed music. Bhandari et al. \cite{bhandari2024motifsphrasesbeyondmodelling} compared different structure representations of symbolic music, and found that music generation still faces the challenges of “modelling nuanced development and variation of themes much like human compositions across extended periods”.

Dai et al. \cite{dai2022missingdeepmusicgeneration, Dai2024InterconnectionsPredictivity} found that generated music lacked structural properties that were in human-created music. These include phrase structure, chord and melody progressions, in which generated music failed to capture the varying patterns of entropy at different parts of the song. In a separate study \cite{dai2020automaticanalysisinfluencehierarchical}, Dai analyzed the relationship between the structure of music and the elements of harmony, melody, and rhythm, and found that patterns in harmony and melody were related to hierarchical structure. Dai's analysis metrics in both studies informs the evaluation methods in our paper.

Existing research \cite{Hadjeres2020Anticipation-RNNGeneration, Roberts2018HierarchicalMusic} has tried to tackle the problem of generating longer sequences of music, and few have found success. Mittal et al. \cite{mittal2021symbolicmusicgenerationdiffusion} explored diffusion architecture to generate long-term melodies, and Zou et al. \cite{zou2021melonsgeneratingmelodylongterm} explored a graphRNN approach.

\subsection{Graph Representation of Music}
One of the core advantage of graph representation of music is the flexibility of design, which can be customized based on the needs of individual music information retrieval (MIR) tasks. The most common one is to construct note-level nodes and construct edges based on note relationships~\cite{Jeong2019GraphPerformance, Karystinaios2022CadenceNetworks, Zhang2023SymbolicEvaluation}. There are also works on bar-level \cite{cosenza2023graphbasedpolyphonicmultitrackmusic} and even phrase-level \cite{zou2021melonsgeneratingmelodylongterm} node designs. 

Research exploring graph representations of music have shown the potential for graph models to perform well in tasks that benefit from learning structural information, such as music classification \cite{dokania2019graphrepresentationlearningaudio,Zhang2023SymbolicEvaluation}, modelling perceptual musical similarity \cite{vahidi:hal-04178191}, and modelling expressive music performances \cite{Jeong2019GraphPerformance}. Furthermore, they provide explainability, and quantify relative importance of various features of the note \cite{karystinaios2024smugexplainframeworksymbolicmusic}. 

\subsection{Music Graph Generation}
% \paragraph{Deep Graph Generation}

% Research in deep graph generation has largely been propelled by molecular science research, where graph generation is particularly useful in representing molecular structure and creating new compounds. There have been various architecture types, such as GraphRNN, GraphVAE, MoFlow, MolGAN, GDSS. \cite{zhu2022surveydeepgraphgeneration} In particular, GraphRNN and GraphVAE are more popular architectures that build on top of RNNs and VAEs. GraphRNN frames graph generation as a next-token prediction problem, and generates nodes and edges sequentially. In contrast, GraphVAE learns to encode and decode graphs as a whole, and generates graphs in a one-shot process.

% In the context of music generation, GraphRNN has the advantage of capturing local dependencies, but might struggle with capturing global structures in music, such as long-term dependencies across different parts of a composition. In contrast, GraphVAE excels at modeling global structures, but could struggle with capturing fine-grained temporal dependencies in music, such as intricate rhythm patterns or subtle melodic nuances.

While there have been research \cite{Kanani2023GraphGeneration, Go2023CombinatorialAnalysis} on graph representations of music for various music tasks, few have explored using graphs for music generation. In particular, two key studies, PopMNet \cite{popmnetWU2020103303}, and MELONS \cite{zou2021melonsgeneratingmelodylongterm} have been pivotal to the study of music generation using graph neural networks (GNNs).

PopMNet introduced a pioneering approach by utilizing a chord-progression and melody structure graph to generate pop music melodies \cite{popmnetWU2020103303}. It successfully captured musical dependencies, but was limited to single-track melodies only. MELONS built on PopMNet by incorporating a bar-level relationship graph and using a GraphRNN with Transformer models to capture long-term structure in melodies \cite{zou2021melonsgeneratingmelodylongterm}. This approach enabled the generation of coherent melodies over extended periods, but is again limited to single-track melodies only, lacking inter-track information such as chords and accompaniment. Additionally, the evaluation relied heavily on subjective methods, and does not give much quantitative measures of its generative ability.

Cosenza et al. \cite{cosenza2023graphbasedpolyphonicmultitrackmusic} introduced Polyphemus, a GraphVAE based model capable of generating polyphonic, multi-instrument music using a chord-level graph. The model generated music that was coherent in rhythm and pitch across multiple tracks, which was an improvement over PopMNet and MELONS, which were melody-only models. The model was highly reproducible, with clear evaluation of the latent space. While Polyphemus was able to capture rhythmic structure in bars, it lacked long-term structure.

Although these studies show the potential for graph-based music generation, challenges such as long-term structure and multi-instrumentality have great room for exploration.
\begin{figure*}
    \centering
    \includegraphics[width=1\linewidth]{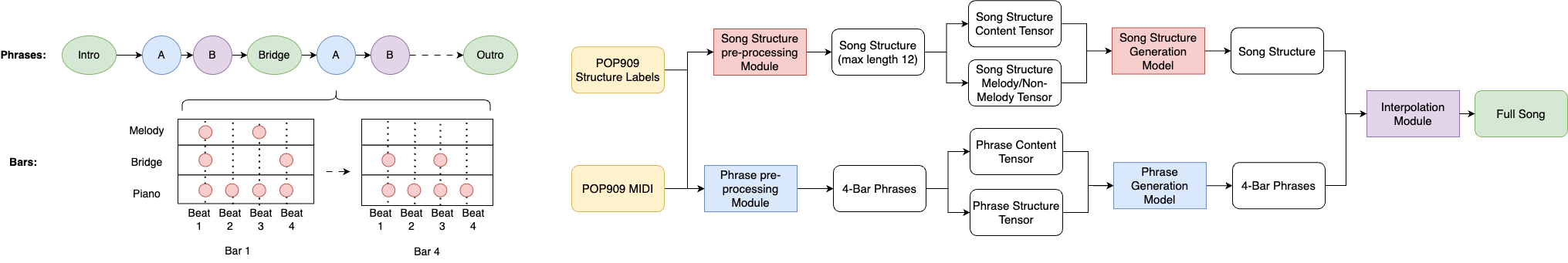}
    \caption{Two-step hierarchical approach: Music is made up of phrases, and each phrase can be sectioned into bars. We train a model for each level of hierarchy.}
    \label{fig:hierarchy_process}
\end{figure*}
\section{Methodology}
To generate polyphonic symbolic music that exhibits coherent long-term structure, we propose GraphMuGen, a two-step approach that involves producing individual musical phrases, and then arranging these phrases into a song structure. GraphMuGen requires two generative models - a \textit{phrase }\textit{generation mode\textit{l},} and a \textit{song structure generation model}. The two models represent two levels of hierarchy of song structure, as illustrated in Figure \ref{fig:hierarchy_process}.

\begin{figure}
    \centering
    \includegraphics[width=0.75\linewidth]{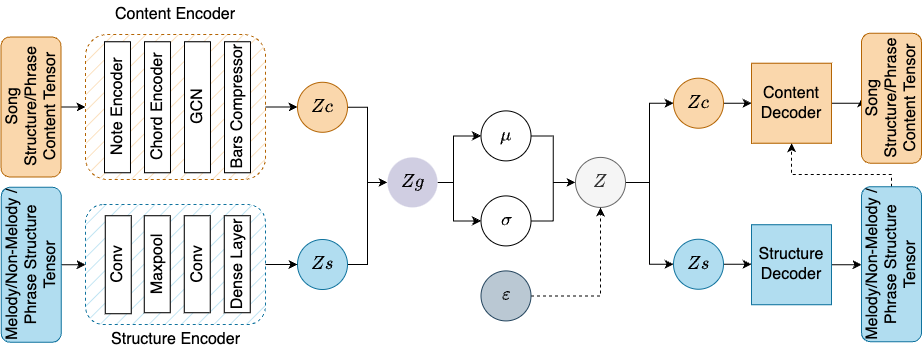}
    \caption{Model VAE Architecture. Both models follow a similar architecture.}
    \label{fig:model_archi}
\end{figure}
\subsection{Dataset and preprocessing}
To generate Chinese pop music, we used POP909 for training both the phrase generation and structure generation models. POP909 consists of 909 Chinese Pop songs in MIDI format, with 3 tracks per song: Melody, Bridge, and Piano accompaniment. We also use structure labels of POP909 \cite{dai2020automaticanalysisinfluencehierarchical}, which are human-labelled, and contains information about the phrase types (i.e. intro, melodic, non-melodic, bridge, outro), and the phrase length in number of bars.

For phrase generation, we filtered the data for 4/4 time songs only. Each song is then split into phrases based on the structure labels. Phrases that are longer than 4-bars are then split further (e.g. 8 bars are split into 2x4 bar tracks), and those with two concurrent empty bars are dropped. 

This method of pre-processing differs from the original method in Polyphemus, which extracted a moving window of N-bars. While a moving window creates a larger and more varied dataset, it loses the structural information within each phrase, since each example could contain bars in between phrases. Our proposed method of splitting by phrases ensures that the model learns structure within a phrase. Arguably, some structural information is still lost when we split 8-bar phrases into 2x4-bar phrases.

For structure generation, the phrase labels were processed to extract the encoded phrase type and phrase lengths. Songs with more than 12 phrases were truncated (i.e. only the first 12 are included). 

For both models, we used a train-val-test split of 70-20-10.

\subsection{Graph Representation}\label{AA}
Each level of structural hierarchy can be represented as a graph. 

\subsubsection{Bar Level}
\begin{figure}
    \centering
    \includegraphics[width=1\linewidth]{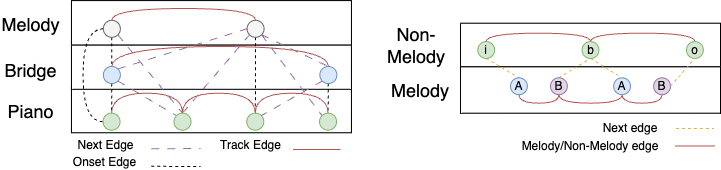}
    \caption{Graph representation of phrase structure of a single bar (left) and song structure (right) in POP909.}
    \label{fig:structure_graph}
\end{figure}

At the bar level, we adopt Polyphemus' method of graph representation (Figure \ref{fig:structure_graph}). Each node in the graph represents a chord at a single time step in each track. A chord is defined as a group of notes played concurrently in a single track. Connecting the nodes are three types of edges, representing different relationships between the nodes: (i) Next Edges connect consecutive nodes across different tracks, (ii) Track edges connect consecutive nodes within the same track, and (iii) Onset edges connect nodes with the same onset time, across all tracks. Each node is represented by an encoding of its Content, which contains information about the pitch and duration of each pitch.

\subsubsection{Phrase Level}
For song structure generation, the graph is at phrase-level. Each node represents a 4-bar phrase (Figure \ref{fig:structure_graph}). There are two types of graph edges: (i) Next edges, which connect consecutive phrases, and (ii) Melody/Non-Melody edges, which connect consecutive melody/non-melody phrases. Similar to the Bar-level nodes, each Phrase-level node is represented by an encoding of its Content, which contains information about the phrase length (i.e. number of bars), and phrase type (e.g.. intro/outro, bridge, etc.).
% \begin{figure}
%     \centering
%     \includegraphics[width=0.75\linewidth]{structure_graph.png}
%     \caption{Graph representation of song structure from Figure \ref{fig:hierarchy}}
%     \label{fig:song_structure_graph}
% \end{figure}
\subsection{Model Architecture}
Both the \textit{phrase generation} and \textit{song structure generation} models use a similar architecture inspired by the Polyphemus model, as shown in Figure \ref{fig:model_archi}. Each model is a VAE which contains two key components: a structure encoder/decoder, and a content encoder/decoder.  

\subsubsection{Polyphemus Architecture}
The content encoder uses a Graph Convolutional Network (GCN), and is responsible for encoding information about pitch and duration, while taking into account the structure of the bars. It works by first encoding the pitch and duration of notes into note embeddings (\(N\)), and then encoding \(N\) into a chord representation.
The chord representation is then passed through the GCN with the bar-level graph representation. The GCN outputs an embedding for each chord, which is compressed into a bar-level content embeddings. They are then concatenated and further compressed to obtain a content embedding (\(Zc\)).

The structure encoder is a Convolutional Neural Network (CNN) with two convolutional layers that takes in a structure tensor representing node activations across time for each track (i.e. whether a node is present or not at each time step). The CNN outputs bar-level structure embeddings, which are concatenated and then compressed to obtain a structure embedding (\(Zs\)).  \(Zc\) and \(Zs\) are concatenated and merged to obtain a graph latent code \(Zg\). Parameters $\mu$ and $\sigma$ are obtained by passing \(Zg\) through a linear layer. A latent vector \(Z\) is then sampled from \(N\)($\mu$,  $\sigma$).

Both decoders largely mirror their respective encoders, using specular layers to decode \(Z\) into content and structure tensors. The structure decoder decodes \(Z\) to output a structure tensor \(S\) indicating node activations. The content decoder decodes \(Z\) while taking into account \(S\), and outputs a content tensor of logits for the pitches and duration of each node.

\subsubsection{Modifications from Polyphemus}
For the phrase generation model, we modified Polyphemus' CNN \cite{cosenza2023graphbasedpolyphonicmultitrackmusic} to work for three tracks instead of four. The drum-related layers in the content encoder that accounted for percussion-pitched notes were removed since POP909 does not have a percussion track. For the song structure generation model, the CNN was further modified to work for two tracks. The number of layers in all models remained the same. 

\subsection{Training}
We trained the phrase generation model with a batch size of 32. Beta \cite{Higgins2016betaVAELB} was set to 0 for the first 5,000 steps of weight updates and increased by 0.001 every subsequent 2,000 steps. The structure generation model had a batch size of 8. Beta was set to 0 for the first 2,000 steps of weight updates and increased by 0.001 every subsequent 800 steps.

Both models follow the hyper-parameter settings in Polyphemus: 8 layers in the encoder and decoder GCN, with a latent dimension of 512. Adam optimizer was used with an initial learning rate of 1e-4, which was decayed exponentially after 8000 gradient updates with a decay factor of 1- 5e-6. Both models were trained for 150 epochs.

\section{Evaluation}

We evaluated our phrase generation model and song structure generation model separately. By sampling from the latent space to generate new phrases and song structures, we compare the distribution with that of the ground truths of human-composed music\footnote{Visualization of metrics and audio samples: https://graphmugen.notion.site/}. 
% Our evaluation is divided into the following parts: First we evaluated the phrase ability to generate new phrases by comparing the distribution and music theory attributes of randomly generated 4-bar phrases to POP909 phrases.
% Song Structure Generation Model: we evaluated its ability to (i) reconstruct song structures in the Test set, and (ii) generate new song structures by comparing the structure distribution of randomly generated song structures to POP909 song structures.

\subsection{Phrase Generation Model}

We randomly generated 100 latent codes, and used the model decoder to generate 4-bar phrases from the latent codes. We measured the quality of the generated phrases by evaluating their (i) similarity to the training distribution, and (ii) ability to reproduce structural nuances.

\textbf{Phrase-level distribution}: e measured the following metrics for each phrase:
\begin{enumerate}
    \item Empty Bar Rate (\(P_{\text{EB}}\)): Defined as the proportion of empty bars to the total number of bars, calculated as \(P_{\text{EB}} = \frac{N_{\text{empty bars}}}{N_{\text{total bars}}}\), where \(N_{\text{empty bars}}\) is the number of empty bars and \(N_{\text{total bars}}\) is the total number of bars in the piece.

    \item Used Pitch Class (\(N_{\text{UPC}}\)): The number of different pitch classes utilized per bar, expressed as \(N_{\text{UPC}}\), where each count reflects the unique pitch classes in a given bar.
\end{enumerate}
\begin{table}[h]
    \centering
    \begin{tabular}{ccccccc}
    \toprule
    & \multicolumn{3}{c}{Empty Bar (EB)} & \multicolumn{3}{c}{Used Pitch Class (UPC)} \\
    \cmidrule(lr){2-4} \cmidrule(lr){5-7}
    & M & B & P & M & B & P \\
    \midrule
    POP909      & 1.01 & 4.69 & 0.39 & 3.33 & 2.62 & 4.68 \\
    GraphMuGen  & 9.85 & 21.8 & 1.25 & 2.03 & 1.39 & 3.33 \\
    \bottomrule
    \vspace{0.1cm}
    \end{tabular}
    \caption{Distribution metrics for POP909 (Ground Truth) and GraphMuGen output.
    % Empty Bar (EB) is the ratio of empty bars (\%). Used Pitch Class (UPC) is the average number of used pitch classes. 
    M: Melody; B: Bridge; P: Piano}
    \label{tab:pitch_metrics}
\end{table}
We compare the metrics for the 100 generated phrases (GraphMuGen) and the pre-processed phrases (POP909) in Table \ref{tab:pitch_metrics}. We observe that our model generates fewer pitch classes and more empty bars compared to POP909 phrases. 
% This is similar to the behaviour observed in reconstruction, where the reconstructed pieces have fewer concurrent pitches, less variation in pitches, and fewer activated time steps as compared to the original piece.
\begin{table}[h]
    \centering
    \begin{tabular}{cccccccc}
    \toprule
    & \multicolumn{4}{c}{Empty Bar (EB)} & \multicolumn{3}{c}{Used Pitch Class (UPC)} \\
    \cmidrule(lr){2-5} \cmidrule(lr){6-8}
    & D & B & S & G & B & S & G \\
    \midrule
    LMD (4-bar)  & 0.79 & 0.64 & 1.1  & 1.5 & 2.27 & 2.91 & 1.91 \\
    Polyph-LMD4  & 0.25 & 0.5  & 6.3  & 5.7 & 2.14 & 1.24 & 1.70 \\
    \bottomrule
    \vspace{0.1cm}
    \end{tabular}
    \caption{Metrics for Polyphemus. 
    % Empty Bar (EB) is the ratio of empty bars (\%). Used Pitch Class (UPC) is the average number of used pitch classes. 
    D: Drums; B: Bass; S: Strings; G: Guitar}
    \label{tab:pitch_metrics_lmd}
\end{table}

For a benchmark comparison, we randomly generated 100 tracks using Polyphemus (Polyph-LMD-4), and compared its metrics against LMD-4 (Table \ref{tab:pitch_metrics_lmd}). We observe that the generated tracks for Drums and Bass have similar empty bars and pitch classes as LMD-4. However, for the Strings and Guitar tracks, the Empty Bar ratio are significantly higher in the generated tracks as compared to the original data. This is similar to what we observed for GraphMuGen in Table \ref{tab:pitch_metrics}. Furthermore, the Used Pitch Class for Strings is also higher in LMD-4 than from the generated tracks. This shows that the model has varying ability to generate pitch classes and activations for different types of tracks. The model performs well for tracks with more consistent rhythm like Drums and Bass, but less so for Strings, Guitar, and Piano parts.

\begin{table*}[h]
    \centering
    \small % Reduce font size
    \setlength{\tabcolsep}{2pt} % Reduce spacing between columns
    \renewcommand{\arraystretch}{1.2} % Adjust vertical spacing between table rows
    \begin{tabular}{@{}ccccccccccccccccccccccccccccccc@{}} % Remove padding
    \toprule
    Bar & \multicolumn{7}{c}{POP909 (Chord)} & \multicolumn{7}{c}{GraphMuGen (Chord)} & \multicolumn{7}{c}{POP909 (Melody)} & \multicolumn{7}{c}{GraphMuGen (Melody)} \\
    \cmidrule(lr){2-8} \cmidrule(lr){9-15} \cmidrule(lr){16-22} \cmidrule(lr){23-29}
    & I & II & III & IV & V & VI & VII & I & II & III & IV & V & VI & VII & 1 & 2 & 3 & 4 & 5 & 6 & 7 & 1 & 2 & 3 & 4 & 5 & 6 & 7 \\
    \midrule
    1 & 32 & 13 & 11 & 9 & 24 & 9 & 2 & 23 & 13 & 17 & 8 & 33 & 4 & 2 & 16 & 16 & 21 & 3 & 18 & 18 & 7 & 11 & 20 & 20 & 7 & 18 & 16 & 8 \\
    2 & 31 & 11 & 14 & 7 & 21 & 12 & 2 & 29 & 15 & 14 & 8 & 32 & 3 & 1 & 15 & 17 & 19 & 3 & 18 & 19 & 7 & 19 & 23 & 10 & 6 & 17 & 16 & 8 \\
    3 & 28 & 12 & 15 & 9 & 18 & 16 & 2 & 26 & 14 & 17 & 7 & 28 & 6 & 2 & 17 & 17 & 18 & 3 & 18 & 19 & 7 & 12 & 20 & 15 & 4 & 18 & 22 & 9 \\
    4 & 27 & 14 & 13 & 11 & 21 & 13 & 2 & 28 & 10 & 18 & 9 & 26 & 9 & 1 & 17 & 19 & 18 & 4 & 17 & 19 & 7 & 21 & 15 & 19 & 5 & 18 & 16 & 7 \\
    \bottomrule
    \end{tabular}
    \caption{Chord and melody frequency probabilities (\%) by phrase position on POP909 and 100 GraphMuGen generated phrases}
    \label{tab:combined_freq_tbl}
\end{table*}

\textbf{Phrase-level music theory attributes}: To address the challenges highlighted by Dai \cite{dai2022missingdeepmusicgeneration}, we adopted the metrics in their analysis to evaluate the model's ability to reproduce phrase-level harmonic trends. We define them as follows:
% \begin{enumerate}
%     \item Chord frequency probability = Count of each chord / number of chords per bar
%     \item Melody pitch class frequency probabilities = Count of each pitch class / number of pitches per bar
%     \item Melody pitch entropy = Shannon entropy of the probability of each each pitch per bar
% \end{enumerate}
\begin{enumerate}
    \item Chord Frequency Probability (\(P_{\text{CF}}\)): Defined for each chord \(i\) as \(P_{\text{CF}, i} = \frac{N_{\text{chord}, i}}{N_{\text{chords, bar}}}\), where \(N_{\text{chord}, i}\) is the count of the \(i\)-th chord and \(N_{\text{chords, bar}}\) is the total number of chords per bar.

    \item Melody Pitch Class Frequency Probability (\(P_{\text{PCF}}\)): Calculated for each pitch class \(j\) as \(P_{\text{PCF}, j} = \frac{N_{\text{pitch class}, j}}{N_{\text{pitches, bar}}}\), where \(N_{\text{pitch class}, j}\) is the count of the \(j\)-th pitch class and \(N_{\text{pitches, bar}}\) is the total number of pitches per bar.

    \item Melody Pitch Entropy (\(H_{\text{MPE}}\)): The Shannon entropy for pitches per bar is computed as \(H_{\text{MPE}} = -\sum_{k=1}^{K} p_{\text{pitch}, k} \log_2(p_{\text{pitch}, k})\), where \(p_{\text{pitch}, k}\) is the probability of the \(k\)-th pitch, and \(K\) is the total number of unique pitches.
\end{enumerate}

The Chord and Pitch classes are based on the inferred Key of the phrase, which were inferred using the 'music21' library \cite{CuthbertA10}.

% For comparability, we re-analyzed the POP909 phrases to ensure that the same inference algorithms were used, and to account for the differences in phrase definition.

In Table \ref{tab:combined_freq_tbl}, we observe that the chord frequency probabilities for GraphMuGen remains mostly similar across all 4 bars. Chords V and I had significantly higher probabilities than other chords, with Chord V having the highest probability in Bar 1, but levelled out with Chord I in bar 4. Comparatively, in POP909, Chord I is is most the frequent chord throughout the phrase. While there is some differences from the chord progressions of POP909, the generated phrases showed similar chord distributions, where chords I and V were most commonly used, and chords VII and IV were more sparsely used.

Comparing melody pitch classes, the generated phrases had more variance in pitch class probabilities (Table \ref{tab:combined_freq_tbl}), while POP909 phrases showed consistent pitch class probabilities throughout the phrase. The generated phrases still captured some of the distributions, where pitch classes 4 and 7 were rarely used.
% \begin{figure}
%     \centering
%     \includegraphics[width=1\linewidth]{melody_pitch_entropy.png}
%     \caption{Comparison of melody pitch entropy by phrase position}
%     \label{fig:mel_pitch_ent}
% \end{figure}

Despite the varying pitch class distribution in our generated phrases, the entropy of melody pitches was lower than that of POP909 
% (Figure \ref{fig:mel_pitch_ent})
. This might be due to our model generating more empty bars than seen in POP909, resulting in sparser pitches. Despite this, the generated phrases captured the structural trend in melody pitch entropies, where the start of phrases have a lower entropy than the rest of the phrase.

\subsection{Song Structure Generation Model}
% We evaluate our model's ability to reconstruct song structure with the following metrics on the Test set:
% \begin{enumerate}
%     \item Melody/Non-Melody accuracy,  precision, and recall
%     \item Type accuracy = Number of correct phrase type / Number of phrases in the original song
%     \item Length accuracy = Number of correct length of phrase / Number of phrases in the original song
% \end{enumerate}
% \begin{table}[h]
%     \centering
%     \begin{tabular}{ccccc} 
%     \toprule
%     & & \multicolumn{3}{c}{Melody/Non-Melody} \\ 
%     \cmidrule{3-5}
%     Type Accuracy & Length Accuracy & Accuracy & Precision & Recall \\ 
%     \midrule
%     88.7 & 83.4 & 91.2 & 92.2 & 88.5 \\ 
%     \bottomrule
%     \end{tabular}
%     \caption{GraphMuGen Song structure reconstruction metrics}
%     \label{tab:reconstruction_structure}
% \end{table}

% As shown in Table \ref{tab:reconstruction_structure}, the model could reconstruct the type, and length of the phrases, and performed well in reconstructing whether the phrase was a Melody or Non-Melody phrase. This means that the reconstructed song structure matched the original song structure well.

Our model is able to reconstruct the type, length and Melody/Non-Melody activations, with 88.7\%. 83.4\%, and 91.2\% accuracies respectively on the Test set.

To evaluate its ability to generate new song structures, we randomly generated 100 latent codes, and used the model decoder to generate song structures from the latent codes. We measured the quality of the generated song structures by comparing its similarity to the training distribution. We adopted the metrics from Dai et al. \cite{dai2020automaticanalysisinfluencehierarchical} in their analysis of structure in POP909. We define them as follows:
\begin{enumerate}
    \item Number of Phrases per Song (\(N_{\text{phrases}}\)): \(N_{\text{phrases}} = \text{total phrases in song}\).
    \item Number of Unique Phrases per Song (\(U_{\text{phrases}}\)): \(U_{\text{phrases}} = \text{unique phrases in song}\).
    \item Frequency Probability of Phrase Length (\(P_{\text{length}}(l)\)): \(P_{\text{length}}(l) = \frac{\text{count of phrases of length } l}{\text{total phrases}}\).
\end{enumerate}

The metrics for POP909 were calculated after truncating each song to have a maximum of 12 phrases.

\begin{table}[h]
    \centering
    \begin{tabular}{ccccccc}
    \toprule
    & \multicolumn{2}{c}{Num  Phrases}
    & \multicolumn{2}{c}{Num Unique Phrases} & \multicolumn{2}{c}{Phrase Length} \\
    \cmidrule(lr){2-3} \cmidrule(lr){4-5}  \cmidrule(lr){6-7}
    & Mean & Std & Mean & Std  & Mean & Std \\
    \midrule
    POP909  & 10.67 & 1.36 & 6.49  & 1.28 & 6.52 & 2.35 \\
    GraphMuGen  & 11.11 & 1.48  & 5.67  & 1.06 & 6.45 & 3.21 \\
    \bottomrule
    \vspace{0.1cm}
    \end{tabular}
    \caption{Metrics for phrase structure distribution}
    \label{tab:phrase_structure}
\end{table}
% \begin{figure}
%     \centering
%     \includegraphics[width=1\linewidth]{num_phrases_per_song.png}
%     \caption{Comparison of the distribution of number of phrases per song}
%     \label{fig:num_phrases_per_song}
% \end{figure}
Table \ref{tab:phrase_structure} shows that the generated song structures (GraphMuGen) have a similar distribution of number of phrases and number of unique phrases. They both have similar mean phrase length, but the generated song structures have more variance in phrase length. In POP909, most phrases were of either 8-bars (35\%) or 4-bars (32\%) long. Our model was able to capture a similar distribution (8-bar: 55\%; 4-bar: 35\%). The phrase types generated were also similarly distributed, with A and B phrases making up most of the phrases for both POP909 (A; 28\% B: 28\%) and our generated phrase types (A; 27\% B: 38\%).

\section{Conclusion and Future Work}

Our research with models like GraphMuGen has shown they accurately replicate the structural properties of training data, but improvements are needed in pitch accuracy and minimizing empty bars. Implementing a temperature threshold to prevent note activation for probabilities under 0.5 could enhance musicality.

During phrase model training, we noted a trade-off between the KL-Divergence (KLD) of the latent embedding and note losses, suggesting difficulties in achieving a well-generalized Normal latent embedding without compromising pitch and duration accuracy. Unlike GraphMuGen, which used only 5,000 examples, the Polyphemus model benefited from a larger dataset from the LakhMIDI database.

To advance our models, expanding the dataset—possibly by augmenting the POP909 phrases with pitch transposition or note variation and incorporating phrase types—could significantly refine outputs. This includes differentiating between introductory and chorus phrases. Our song structure model currently captures basic phrase attributes like length and type but could improve by adding more complex phrase details such as pitch and structure. Enhancing this aspect would allow the model to understand inter and intra-phrase relationships better, leading to more accurate content tensor generation. Furthermore, applying a heuristic in full song generation to select only tracks of the same key and removing non-melodic sections could enhance the consistency and quality of the music produced.

The efficacy of GraphMuGen in generating polyphonic symbolic music with complex, long-term structure highlights the potential of graph representations in music generation. Despite some challenges, both models retained most structural properties observed in the training data, marking a promising step towards generating symbolic music that mirrors the structural complexities found in human-created music.

\bibliographystyle{abbrvnat}
\bibliography{references, ref}

\vspace{12pt}

\end{document}